\documentstyle[aaspptwo,psfig]{article}

\def\etal{et al.}
\def\hhp{m21}
\def\hb{m28}
\def\cmd{{\rm CMD}}
\def\cmds{{\rm CMDs}}

\def\hhpfwha{$0.6''$}
\def\hhpfwhb{$1.4''$}
\def\hbfwhm{$0.6''$}

\def\PSF{PSF}
\def\ccd{CCD}
\def\feh{{\rm[Fe/H]}}

\begin{document}

\title{ A $BVI$ Photometric Study of the Near-Galactic Center
Globular Cluster NGC 6517 (C1759-089) }
\author{ JJ Kavelaars,  David A. Hanes\altaffilmark{1}\\  Terry J. Bridges \altaffilmark{1,2}}
\affil{Physics Department, Queen's University, Kingston, ON, Canada \\
Electronic Mail: jj@astro.queensu.ca, hanes@astro.queensu.ca, \\
 tjb@mail.ast.cam.ac.uk }
\author{William E. Harris\altaffilmark{1}}
\affil{Department of Physics and Astronomy, McMaster University, Hamilton, ON, Canada \\
Electronic Mail: harris@physics.mcmaster.ca }

\altaffiltext{1}{Visiting Astronomer, Canada France Hawaii Telescope,
operated by the National Research Council of Canada, le Centre National de la
Recherche Scientifique de France, and the University of Hawaii. }
\altaffiltext{2}{present address: Royal Greenwich Observatory, Madingley Road,
Cambridge, England }

\begin{abstract}

We present the results of a $BVI$ photometric study of the globular
cluster NGC 6517 using CCD images obtained at the
prime focus of the CFHT.  {}From the resultant color-magnitude diagram,
we infer that the cluster is differentially reddened and develop a
method for determining the amplitude of this effect, which amounts to
\( \sim 0.4 \) mag in \( (V-I) \) across our $\sim 4$ arc-min field.  {}From the
corrected $V_0,$ $(V-I)_0$ color-magnitude diagram, we derive the metallicity
and distance modulus of the cluster.  Our values for the mean foreground
reddening ($E(B-V) = 1.10 \pm 0.10$) and metallicity (\feh\ = -1.58
$\pm$ 0.05) agree to within uncertainties with previous determinations.  
Our apparent distance modulus of
18.3 $\pm$ 0.2, however,  is larger than the sole
previous determination. When the distance modulus is corrected for absorption,
assuming that $R_V = 3.8 $ for this line of sight, we find that
NGC 6517 is within 3 kpc of the Galactic center, making it a bulge member.
This makes NGC 6517 only the fourth metal poor cluster in the bulge for which 
an accurate CMD is available. 

\end{abstract}

\section{Introduction \label{ch:intro}}

  How the Galaxy formed is one of the most debated questions in astronomy
  today.  As a result of an improved understanding of stellar evolution and
  better observations of the stellar populations of the Galaxy,
  early models of galaxy formation,
  such as the Eggen, Lynden-Bell, \& Sandage (1962) picture,
  \nocite{Eggen62} have been amended to the
  more recent versions of Searle \& Zinn (1978)
  \nocite{Searle78} and
  Lin \& Murray (1991). \nocite{Lin91}
  Major constraints on these theories are derived {}from
  observations of the globular cluster system (GCS) of the Galaxy.

The GCS has two principal components
(\cite{Morgan59}; \cite{Harris76}; \cite{Searle78} and \cite{Zinn85}).
The {\em disk} subsystem is
somewhat flattened, with typical rotational velocities about the
galactic center of $\sim 170$ km/s; the disk clusters are metal-richer
than $[Fe/H] = -0.8$ (\cite{Armandroff89}).  The {\em halo} subsystem of
clusters rotates more slowly, at $V_{rot} \approx 45$ km/s, and
consists of the metal-poor clusters, $ [Fe/H] < -0.8$.  The halo
subsystem is sometimes further subdivided into an inner halo, with
$R_{GC} < 7 kpc$, and an outer halo, for which $R_{GC} > 7 kpc$.  For a
complete review of the globular cluster system of the Galaxy see
\cite{Zinn85} and \cite{Armandroff89}, and references therein.

A thorough study of the globular cluster system is essential, therefore, if
any conclusions about the formation of the Galaxy are safely to be
drawn.  Prior to 1990, however, no accurate \cmds\ had been published
for halo clusters within $\sim 3$ kpc of the galactic center
(\cite{Janes91}).  It has been suggested, on the basis of rough distance
estimates, that NGC~6517 is such a cluster:  see Section 2.  Moreover, NGC
6517 is
known to be of intermediate-to-low metallicity,
and it is consequently an especially interesting object for
closer scrutiny as an inner halo object.  Only recently, however, with the
combination of charge coupled devices (CCDs), specialized image
processing packages, and excellent seeing, have the necessary studies of
extremely crowded objects like NGC 6517 become possible.  Without
observations of such inner halo clusters our understanding of the
globular cluster system will remain incomplete.

This paper will take the following form:  In Section~2, previous
observations of NGC 6517 are presented, along with a review of
determinations of the cluster's distance and metallicity.  In Section~3
we describe our own observations, the initial stages of the data
reduction, and our calibration.  In Section~3.3, the particular question
of reddening is addressed.  It is apparent {}from the width of the \cmd\
that NGC 6517 is differentially reddened.  Further evidence for this
conclusion, as well as a rough modeling of the effect, are presented and
quantified.  In Section~4, we present the corrected \cmd\ for NGC 6517,
derive the distance and metallicity of the cluster, and compare our
results to earlier determinations.  Finally, in Section 5, we summarize
our principal findings.

\section{Previous Observations \label{ch:oldobs}}

In Table~\ref{tab:N6517}
 we list the heliocentric position, the galactocentric distance, and some
of the characteristics of NGC 6517. All the values are {}from Webbink's (1985) 
tabulation, \nocite{Webbink85}
except the radial velocity, which is {}from Zinn (1985) \nocite{Zinn85}.
The quoted galactocentric distance assumes $R_o = 8.8$ kpc and
$V({\rm HB})$ = 18.0; the present study leads to revision of this
value (see Section 5).  Figure~\ref{fig:plate} is a
reproduction of a CCD frame of NGC 6517 taken through the $V$ filter.
At the prime focus of the CFHT, the 640 x 1024 RCA4 CCD which we used has
a scale of 0.22 arc-sec/pixel and a field  of 2.35 x 3.76 arc min.
The readnoise of the chip was $\sim 64 e^-.$

\begin{table}
\caption{Properties of NGC 6517}
\begin{center}
\begin{tabular}{ll}
\tableline
\tableline
\\
$\alpha_{1950}$ & 17 59 06 \\
$\delta_{1950}$ & -08 57.6  \\
$\ell$             & 019.225    \\
$b$             & +06.762   \\
$R_{GC} $       & $\sim 3.8 kpc$ \\
$V_{radial}$       & $-47 km/s$ \\
$M_{V} $&       $-7.10$ \\
\\
\tableline
\tableline
\end{tabular}
\end{center}
\label{tab:N6517}
\end{table}

\begin{figure}[hbt]
\caption[$V$ filter image of NGC 6517]{A 60s $V$ image of NGC 6517 taken at
the CFHT.  North is to the top, and East to the left;
the field is 2.35 x 3.76 arc-min.}
\label{fig:plate}
\end{figure}

We now briefly review
earlier photometric and spectroscopic studies of the cluster.


Integrated colors for NGC 6517 have been obtained by Kron \& Mayall
(1960) \nocite{Kron60}
in the $BPI$ system, and by Harris \& van den Bergh (1974), \nocite{Harris74}
 Zajtseva et al. (1974) \nocite{Zajtseva74} and Racine (1975) \nocite{Racine75}
 in the $UBV$ system, with the results as summarized in Table 2.

\begin{table*}[htb]
\caption{Previous observations of NGC 6517.}
\begin{center}
\begin{tabular}{rcccccc}
\tableline
\tableline
\\
Author(s) & $(B-V)$ & $(U-B)$ & $E(B-V)$ & $(m-M)_{\rm V}$ & [Fe/H] \\
\tableline
\cite{Kron60}   & 1.75 &     &           &   &  \\
\cite{Harris74} & 1.80 & 0.81 & $1.14\tablenotemark{a}$ &   & \\
\cite{Zajtseva74}& 1.79 & 0.98 & $1.07\tablenotemark{a}$ &   & \\
\cite{Racine75}  & 1.75 &0.88 & $1.05\tablenotemark{a}$ &   & \\
\cite{Harris75}  &     &     &  1.0       & $18.1\tablenotemark{b} $ &  \\
\cite{Harris80}  &      &     &             & $17.4\tablenotemark{c}$ & \\
\cite{Bica83}    &1.76 &      &   1.11      &   &$-2.18 \pm 0.4$ \\
\cite{Zinn84}    &      &      &             & &  $-1.34 \pm 0.15$ & \\
\cite{Zinn85}    &     &     &   1.08      &  $17.09\tablenotemark{d}$ & \\
\cite{Armandroff89}  &  &      &             &   17.35$\tablenotemark{e}$ & \\
\\
\tableline
\tableline
\end{tabular}
\end{center}
 
\centerline{ Notes to Table~\ref{tab:prevobs}}
\tablenotetext{a}{Reddening determined using
Racine's (1973) intrinsic relation for globular cluster $(B-V)$, $(U-B)$ colors}
 
\tablenotetext{b}{$M_V({\rm HB}) = 0.6 $ (\cite{Sandage70}) }

\tablenotetext{c}{ Distance modulus re-estimated by Harris following
reexamination of the original plate material (Harris 1975).}
 
\tablenotetext{d}{$M_V({\rm HB}) = 0.35  * ([Fe/H] - (-1.66)) + 0.8$ (\cite{Sandage82}
) }
 
\tablenotetext{e}{$M_V({\rm HB}) = 0.20  * ([Fe/H] - (-2.3)) +  0.46$
                                (\cite{Lee87})  }

\label{tab:prevobs}
\end{table*}


The only previous \cmd\ of NGC 6517 was constructed by Harris
(1975)\nocite{Harris75} as part of a $B$,$V$ photographic
investigation of 12 southern galactic globulars, using the 0.6-m
University of Toronto telescope at Las Campanas Observatory.  Plate
limits of $V\approx18$ and $B\approx19$ were reached.  In his
description of the cluster, Harris states ``As well as being small,
distant and highly reddened, this difficult object is compact...''.  As
a result of the high background, the severe crowding,
and the compactness of the cluster, the $(B - V)$ colors were
not reliable below $V\approx17$.  These very shallow limits made a
direct determination of the distance to NGC 6517 {}from its horizontal
branch (HB) impossible.

However, Harris inferred a horizontal branch apparent magnitude of $V =
18.7$ for the cluster by inter-comparing its giant branch (GB) with those
of other clusters of the same spectral type and relying on previous
determinations of GB height as a function of spectral type 
(\cite{Sandage60}).


Bica \& Pastoriza (1983) \nocite{Bica83} included NGC 6517 in a survey
of 91 galactic globulars observed both with DDO and with $UBV$ filters.
Although there were many discrepancies in the metal-poor ($\feh\ <
-1.5$) range, they found that their integrated DDO colors correlated
with the existing metallicity determinations for many of
the clusters in their survey; consequently, they were able to use the
DDO colors to predict reddenings in $UBV$.  Their determinations of
\feh, $E(B-V)$, and $(B-V)$ for NGC 6517 are given in
Table~\ref{tab:prevobs}.

Zinn \& West (1984) \nocite{Zinn84} included NGC 6517 in a survey of
integrated cluster spectra.  Their method of determining metallicities
relies upon the size of the $Q_{39}$ index (\cite{Zinn80}), a measure of
the strength of the Ca H and K features in a globular cluster's
integrated spectrum.  Zinn (1985) \nocite{Zinn85} also found that $Q_{39}$
correlated well with the integrated $(B-V)_0$ colors, and used
previous determinations of the color of NGC 6517 to derive
the reddening (Table~\ref{tab:prevobs}).


An examination of Table~\ref{tab:prevobs} reveals that the estimates of
reddening for NGC 6517 are broadly consistent between authors.  A
straight mean of those reddening values in Table~\ref{tab:prevobs} that
were derived using the intrinsic color-color relation of Racine (1973)
is $E(B-V) = 1.09$. This value is in excellent agreement with that
found spectroscopically by Zinn (1985).  \nocite{Zinn85} In contrast,
there are only two determinations of [Fe/H], and there is a large
discrepancy between them.

Table~\ref{tab:prevobs} lists various determinations of the apparent
distance modulus in $V,$ but it must be emphasized that, even though
these estimates differ, {\em all} of the values rely on Harris's (1975)
original \cmd\ for the cluster.  In that paper, Harris estimated the tip
of the giant branch (GB) to lie at $V = 16$, and took the HB to lie
$\Delta V=2.7$ mag fainter; in this way, he deduced $V({\rm HB}) = 18.7$.
In a later work, however, \nocite{Harris80} Harris (1980) revised his
estimate of $V({\rm HB})$ {}from 18.7 to 18.0, upon a re-examination of the
cluster's red giant branch (RGB).  Using this
revised $V({\rm HB}),$ Zinn (1984) \nocite{Zinn84} and Armandroff (1989)
\nocite{Armandroff89} calculated the distance modulus of NGC 6517
on the basis of two different relations between the absolute magnitude of the
HB and the cluster metallicity (see the footnotes to
Table~\ref{tab:prevobs}).  We shall return to this point when we carry out
our own distance determination in section 4.1.

\section{New Observations and Analysis
\label{ch:newobs}}

Our observations were carried out as part of two separate runs in May
1990 at the prime focus of the 3.6-m Canada-France-Hawaii Telescope
(CFHT).  These observations were made with the RCA4 \ccd\ and a $BVI$
filter set designed to match the standard system.  Standard stars {}from those
tabulated by Landolt (1983) \nocite{Landolt83} and Christian \etal\
(1985) \nocite{Christian85} were observed each night to provide
calibration.

Historically, globular cluster \cmds\ have been constructed in the $(B - V),$
$V$ plane.  However, NGC 6517 is a heavily reddened object
and $B$ filter exposures would be uneconomical
as they would
require very long exposure times.  For this reason, we decided
to use the $I$ filter.
Indeed recent work (\cite{DaCosta90}) has
has shown that the $(V - I)$ color is very useful in determining
cluster metallicities.
Nevertheless, in addition to our $V$ and $I$
exposures, we did secure a series of short $B$ images in order
that an estimate of
the reddening could be made {}from a color-color diagram.

\begin{table}
\caption[Observing Log]{Observing log for program field, both nights.}
\begin{center}
\begin{tabular}{rccccc}\tableline\tableline
\\
Night & Exposure(s) &Filter & Seeing    & Airmass \\ \tableline
\hhp  & 3x100       &     $V$  & \hhpfwha &  1.207    \\
\hhp  & 6x100       &    $I$  &\hhpfwha  & 1.223     \\
\hhp  & 6x100       &     $V$  & \hhpfwhb  & 1.258     \\
\hhp  & 3x100       &    $I$  &\hhpfwhb  & 1.301     \\
\hb   &  60        &     $B$  & \hbfwhm   & 1.205    \\
\hb   & 10x100      &    $B$ & \hbfwhm   & 1.210    \\
\hb   & 60          &    $V$  & \hbfwhm   &1.328    \\
\hb   & 60          &    $I$  & \hbfwhm   &1.339    \\
\\
\tableline
\tableline
\end{tabular}
\end{center}
\label{tab:obssum}
\end{table}

In Table~\ref{tab:obssum} we summarize the important features
of the observing sessions.
On the night of May 21 (m21 hereafter), the seeing during our
observations of NGC 6517 was
extremely good to begin with, \hhpfwha, but worsened steadily:  by the
time the second set of $V$ images was being taken it had
deteriorated to \hhpfwhb.  The effects of crowding make the attainable
precision very sensitive to the seeing, so we chose simply to reject the
last six frames in $V$ and the last three in $I$ as they provided
negligible improvement to the signal-to-noise.  The data {}from May 28
(m28) consists of 11 $B$ frames, one in $V,$ and one in $I,$ with the
last two taken to test the consistency of the photometry {}from the
\hhp\ night.

\subsection{Reduction of Standards}

The standard star observations were reduced with the DAOPHOT
package in IRAF\footnote{IRAF: the Image Reduction and Analysis Facility is
distributed by the National Optical Astronomy Observatories, which is
operated by the Association of Universities for Research in Astronomy, Inc.
(AURA) under cooperative agreement with the National Science Foundation~(NSF).}.
The instrumental magnitudes were transformed to the
standard system using equations of the form:

\begin{equation}
	m - M =  \alpha*C + \kappa*X + \gamma
\end{equation}

\begin{equation}
	c = \alpha*C + \kappa*X + \gamma
\end{equation}

where $m,M,c,C$ are respectively the instrumental and standard
magnitudes and colors.  Unfortunately, there were an insufficient number
of observations of standard stars at large airmass to allow an accurate 
determination  of the
extinction term.  For this reason we adopted for the extinction
coefficients an average of values typical of the CFHT site in previous
observing runs.  The adopted and derived coefficients -~along with the
``notional" values for the CFHT~- are given in
Table~\ref{tab:standcoef}.  The random scatter in the photometry of the
standards is typically $\pm 0.04$ magnitude.

\begin{table}
\caption[Transformation Coefficients]{Coefficients for the transformation
of instrumental $bvi$ magnitudes to the standard system.
Also given are the CFHT notional values.}
\begin{center}
\begin{tabular}{llll}\tableline\tableline 
\\
Coeff.                  &  \hhp         &  \hb          &  CFHT    \\ \tableline
$\alpha_{I}$            &  0.029         &  0.030       &  \\
$\kappa$                 &  \nodata      & \nodata       &  0.091    \\
$\gamma$                &  1.247        &  1.301        &  \\ \tableline
$\alpha_{(V-I)}$        &  0.050        &  0.027        &  \\
$\kappa$                 &  \nodata      & \nodata       &  0.164    \\
$\gamma$                &  0.505        &  0.522        &  \\
$\alpha_{(B-V)}$        &  \nodata          &  0.852    &  \\
$\kappa$                 &  \nodata      & \nodata       &  0.07    \\
$\gamma$                &  \nodata      &  0.126        &  \\ \tableline
\\
\tableline
\tableline
\end{tabular}
\end{center}
\label{tab:standcoef}
\end{table}

\subsection{The Program Frames}

All of our exposures were found to be read-noise dominated, which
suggests that they should be averaged;
 this was done for the $B$ frames {}from the \hb\ night.  For
the \hhp\ night, however, variations in the point spread function (\PSF)
between the frames would have made this averaging unreliable.  For this
reason, the \hhp\ frames in best seeing were simply registered and
co-added.  All frames were then reduced using the DAOPHOT \PSF\ fitting
package in IRAF.\footnote{Our complete star list (in APPHOT
format) along with copies of our combined $B,$ $V$ and $I$
 images are available via
anonymous ftp to Astro.QueensU.CA.}
 Finally, the aperture corrections needed to tie the
zero-point of the \PSF\ magnitudes to the standard system were
determined to a precision of  $\pm0.03$ magnitudes.

In the NGC 6517 photometry, we find no important zero-point differences 
between the $V$ and $I$ scales on the two nights, and no scale 
errors as a function of magnitude.  In combining the photometry {}from
the two nights we reduced the \hhp\ $V$ and $I$ photometry to the \hb\ scales,
for it was on the later night that the photometric standards yielded a
better fit.

\subsection{Reddening \label{sec:red}}

Our observations are tied to the standard system of Cousins,
in which color excesses are related by (\cite{Dean78})

\begin{eqnarray*}
\frac{E(V-I)}{E(B-V)} & = & [1.00 + 0.06*(B-V)_0 \\
                      &   & + 0.014*E(B-V)]
\end{eqnarray*}

For typical RGB stars, $(B-V)_0 = 1.2$, and for NGC 6517 in
particular earlier estimates of the mean foreground reddening yield
$E(B-V) \approx 1.1$ (Table~\ref{tab:prevobs}); this implies
$E(V-I)/E(B-V) = 1.36 $ for the RGB stars in NGC 6517.
  In order to translate the
reddening into an absorption and thus determine the absolute distance
modulus it is necessary to make some assumption about the value of the ratio
of total to selective absorption ( $R_V = A(V)/E(B-V) $).  The correct 
value of $R_V$ is dependent on the nature of the material which is absorbing
the light. For the diffuse interstellar medium $R_V=3.1$ is reasonable but
 within	
areas of dense clouds a value of $R_V=5$ is more appropriate
 (\cite{Cardelli89}).

Before a reconsideration of the total mean absorption in the direction of
NGC 6517, however, we first consider the possibility that the cluster is
{\em differentially} reddened.

\subsubsection{Differential Reddening}

\begin{figure}[hbt]
\hspace*{\fill}\psfig{file=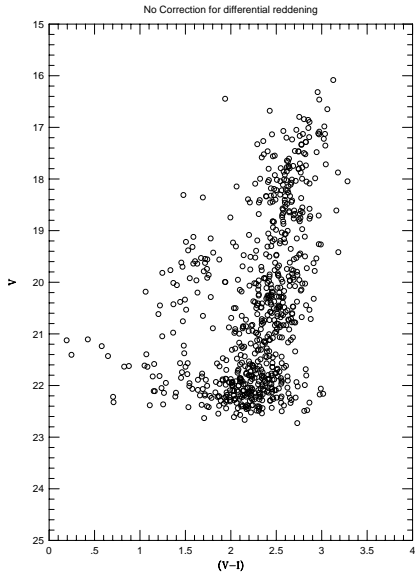,height=3.5in,angle=0}\hspace*{\fill}
\caption[The raw CMD]{The raw \cmd.
Stars with $V < 18 $ are {}from \hb;
all other stars are {}from \hhp.  No stars within $25''$ of the cluster
center or with $\sigma_{(V-I)} > 0.10 $ are included.
Note that the lack of points in the region $18.7 < V < 19.1$ is caused by 
an undersampling of the \cmd\ in this region.  Stars which are brighter than
$V \approx 19 $ are saturated on the \hhp\ frames while stars fainter
than this have less reliable photometry on the \hb\ night.  These effects
conspire to give an underpopulation in this area of the \cmd.  }
\label{fig:rawcmd}
\end{figure}

In Figure~\ref{fig:rawcmd} we present a \cmd\ in $V,$ $V-I$, which shows at
once that the GB is very broad in comparison with typical cluster GBs
(see for example \cite{Hesser88}).  There are two possible explanations:
first, the cluster could have a large spread in metallicity;
or second, the cluster could be differentially reddened.  A spread in
metallicity of 0.8 dex would be required to explain the width of the
GB\footnote{ Based on the $(V-I)_{0,-3}$ to \feh\ relation
(\cite{DaCosta90}).}; this seems implausible.  Only a few clusters
display a spread of metallicity in their member stars, the best-known
example being $\omega\ Cen$ (\cite{Bell81}); for this cluster the spread is
$\Delta \feh\ \approx 0.5 $. Moreover, more than half of the foreground
reddening of NGC 6517 is caused by a single cloud complex, with the remaining
reddening caused by a dense OH cloud (\cite{Sandell87}).  Given the
cluster's proximity to the galactic center and the evident clumpiness of the
foreground material, spatially variable
absorption seems likely.  Indeed we shall now show via a direct test
that this is the case.

In our determination of the presence of differential reddening which follows, we
adopted values of  $R_V$ ranging {}from $R_V=2.5$ to $R_V=5$. The value of
$R_V$ = 3.8 provided the best improvement in the test which we now describe.

In order to determine the nature of this differential reddening
we examined the way in which the mean $V-I$ color of the RGB stars varies with
spatial position on the sky.
To accomplish this, we
segmented the image into a 3 by 3 grid, with  stars less than $25''$ {}from the
center of the cluster excluded. The mean color for
all the RGB stars brighter than $V = 18.5$ was calculated; this revealed that
stars in the northeast quadrant of the frame are bluer in the mean than stars
in the southwest.
This is consistent with
the presence of a Lynds dark cloud to the southwest of the cluster
(\cite{Sandell87}).
The exact nature of the variation in the reddening is most likely very
complex. However, we chose to approximate it as a linear gradient across
the field.
In order to quantify this
variation, the following procedure was adopted:

\newcounter{listcont}
\begin{list}%
{\arabic{listcont}}{\usecounter{listcont}\setlength{\rightmargin}{\leftmargin}}

\item A mean RGB sequence was fit to the ensemble of points in the $V,$ 
$(V-I)$ plane. The offsets of points {}from this sequence, along reddening
trajectories,
 were determined for stars spanning a limited range of the RGB.
 (We will refer to these offset values as $\delta$.)

\item A rectangular coordinate grid was set up across the image in an arbitrary
orientation.

\item The $\delta$ values were fit as a linear function of x position.

\item {}From this fit, positionally dependent reddening corrections 
were calculated
for and applied to all stars in the field, and a
revised \cmd\ was plotted.  Within the \cmd\, the width of the giant branch
(i.e. the spread of points about the mean color as a function of V magnitude)
was calculated.

\item The process was repeated with systematic iterations through various
position angles of our coordinate grid.  The minimum width of the RGB is
expected to come, of course, when the x-axis lies along the true gradient of
the reddening.
\end{list}

\begin{figure}[hbt]
\hspace*{\fill}{\psfig{file=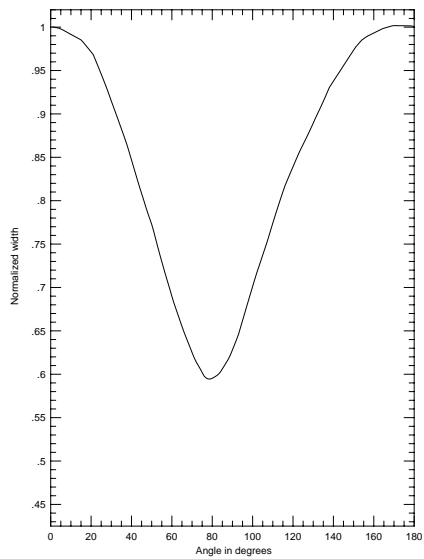,height=3.5in,angle=0}}\hspace*{\fill}
\caption[Giant Branch Width]{
Normalized width of the giant branch as a function of the orientation
of the assumed linear gradient in the differential reddening, as described in
the text.}
\label{fig:gbwidths}
\end{figure}

In Figure~\ref{fig:gbwidths}, we show the results of this experiment.  There
is a clear minimum near an angle of $70^o$ on our coordinate system (which
corresponds to $20^o$ {\em west} of {\em north} on the sky), with an
amplitude of $E(V-I) \sim 0.4$ {}from corner to corner.

\begin{figure}[hbt]
\hspace*{\fill}{\psfig{file=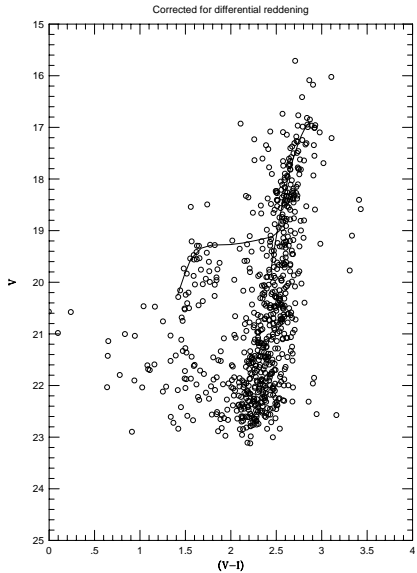,height=3.5in,angle=0}}\hspace*{\fill}
\caption[The Final CMD]{The \cmd\ for NGC 6517 corrected for the effects of
variable interstellar obscuration.
Stars with $V < 18.7 $ are {}from \hb;
all other stars are {}from \hhp.  No stars within $25''$ of the cluster
center or with $\sigma_{(V-I)} > 0.10 $ are included.
The RGB and HB are plotted as fiducial lines chosen to reflect the
most likely positions of the unobscured RGB and HB. }
\label{fig:final}
\end{figure}

Figure~\ref{fig:final} shows the \cmd\ after the data were corrected
for the differential reddening.  In the figure, the data have been
corrected for the {\em average differential} effect:  that is, the scales
are as would be observed at the cluster center.

  It is worth
emphasizing that the experiment just described relies on changes in the width
of the RGB between  $17 < V < 18$ 
 but that the distribution of points in the
corrected figure is manifestly tightened over a much larger range; this
speaks well for the correctness of the model.  Moreover, it is clear that
differential reddening rather than a metallicity spread must be responsible
for the scatter of data in the raw \cmd.

\subsubsection{A Reconsideration of the Global Reddening}

\begin{figure}[hbt]
\hspace*{\fill}{\psfig{file=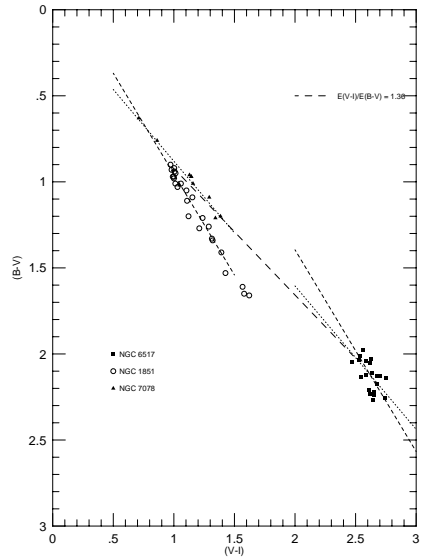,height=3.5in,angle=0}}\hspace*{\fill}
\caption[Two-color Diagram] {
Two-color plot of the GB of NGC 6517 compared to the dereddened
GB of NGC 1851 and NGC 7078.
A line has been fitted to both data sets and both lines
shifted along the reddening trajectory (shown as a dashed line) to best
fit the NGC 6517 data.}
\label{fig:colcol}
\end{figure}

Figure~\ref{fig:colcol} shows the $(B-V),$ $(V-I)$ color-color plot after the
correction for differential reddening.  The {\em mean} foreground
reddening for NGC 6517 was determined {}from a comparison of the RGB
sequence in this plot to those for the clusters NGC 1851 and NGC 7078,
data for which are also shown in the figure (\cite{DaCosta90} and
\cite{Stetson81}).
NGC 1851 and NGC 7078 were used as comparators 
 because the reddenings of these clusters are
well determined and their metallicities, \feh\ =-1.29 and \feh\ =--2.17 
respectively (\cite{DaCosta90}),
 bracket the previous determinations of the metallicity of
NGC 6517.

Straight lines were fit through the RGB of the two comparison clusters and
lines with the same slopes were fit, in turn,
to the NGC 6517 RGB.  The zero-point offsets along  the
reddening trajectories were determined.
The global reddening was then taken as the average of the reddenings
implied by these offsets.
{}From this procedure we determine that for NGC 6517
$E(B-V) = 1.05 \pm 0.15 $ implying that $A_V =  4.0 \pm 0.6 $.

\begin{figure}
\hspace*{\fill}{\psfig{file=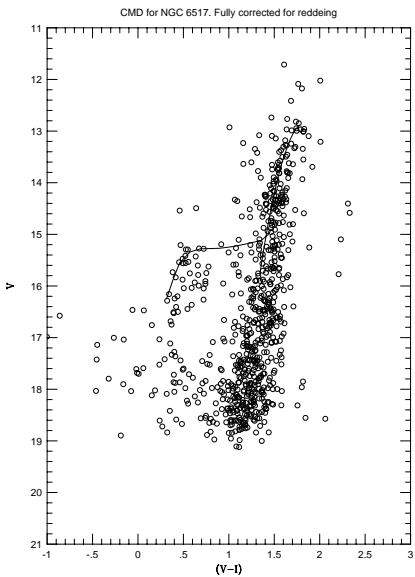,height=3.5in,angle=0}}\hspace*{\fill}
\caption{The CMD for NGC 6517 fully corrected for obscuration.}
\label{fig:abs_cmd}
\end{figure}

Figure~\ref{fig:abs_cmd} gives the final fully corrected CMD for NGC 6517.
The superposed fiducial is a hand-drawn line to best represent the position of 
the cluster's RGB.
The apparent width of the RGB cannot be attributed to photometric uncertainties 
alone and is likely due to residual differential
 obscuration, which is presumed to be patchy on small scales.

\section{Interpretation \label{ch:results}}

In recent years,
refinements in theory have made it
possible to follow the stellar evolution of globular cluster
stars {}from the main sequence to the end of the
HB phase, although the evolution {}from the GB to the HB
is still only phenomenological.
These theoretical models make it possible to analyse the \cmd\
for abundance ratio and age effects (see for example
\cite{VandenBerg91}; \cite{Lee87} and \cite{VandenBerg90}).
However, there remains a great deal of uncertainty
in the zero point of the age scales and the precise evolution
of the member stars in these models.  In what follows, therefore, we prefer
to compare the \cmd\ of NGC 6517 to those of several clusters of similar
composition, thereby avoiding some of the problems of scale calibration.

\subsection{The Metallicity}

Recent work has shown that when the distance modulus and
reddening are known it is possible to determine the metallicity of a cluster
using a cubic relation between the $(V-I)$ color of the RGB (read at a fixed I
absolute magnitude) and cluster metallicity (\cite{DaCosta90}).
The width of our RGB for NGC 6517 makes accurate determinations of the 
fiducial sequence and reddening difficult.
For this reason we do not employ the method developed by Da Costa \& Armandroff (1990). 
We instead rely on the slope, S, of the RGB as our indicator of metallicity.
  This method has
the advantage of being independent of reddening and providing a simple one-step
estimate of the metallicity.

Originally the S parameter (\cite{Hartwick68}) was established in the 
$V,$ $(B-V)$
system. In order to use it here we have calibrated the $S_{(V-I)} $, [Fe/H] 
relation using the clusters tabulated in Da Costa \& Armandroff (1990).  
We define $S_{(V-I)} $ as the slope of the RGB {}from the level of
the HB to a point 2.3 magnitudes higher.
The results of our calibration are given in Table~\ref{tab:fehS}.
An unweighted linear regression
yields $$ {\rm[Fe/H]} = 6.78\times 10^{-2} - 2.561\times 10^{-1} 
\times S_{(V-I)} \pm 0.06.$$

\begin{table}[htb]
\caption[The S Parameter]{S parameter relation for $(V-I),$ all
metallicities are {}from Da Costa \& Armandroff (1990).}
\begin{center}
\begin{tabular}{lll}\tableline\tableline
\\
Cluster & [Fe/H] & $ S_{(V-I)} $ \\ \tableline
NGC 7078 & -2.17 & 8.31 \\
NGC 6397 & -1.91 & 7.72 \\
NGC 7089 & -1.58 & 6.93 \\
NGC 6752 & -1.54 & 6.30 \\
NGC 1851 & -1.29 & 4.93 \\
NGC 104  & -0.71 & 3.04 \\
\\
\tableline
\tableline
\end{tabular}
\label{tab:fehS}
\end{center}
\end{table}

{}From Figure~\ref{fig:abs_cmd} the color of the RGB 2.3 magnitudes 
above the HB is $(V-I) = 2.9 \pm 0.1$
and $(V-I)_{\rm HB} = 2.55 \pm 0.1$; thus $S_{(V-I)} = 6.57 \pm 0.15 $, and 
we deduce that [Fe/H] $= -1.62 \pm .06$.

Sarajedini (1994) \nocite{Sarajedini94}
 has shown that it is  possible to determine both 
[Fe/H] and $E(V-I)$ simultaneously {}from a $V,$ $(V-I)$ \cmd.
Using this technique we find that [Fe/H] $= -1.55 \pm 0.15$ and 
$E(V-I) = 1.55 \pm 0.15 $ with the uncertainty
resulting {}from the width of the RGB and the uncertainty in the position of the
HB.  These values agree, within the quoted uncertainties, with those determined
using the slope of the RGB and the offset of the RGB in the two-color plot. 

The use of these two methods provides an internal consistency check on the
RGB as they are both calibrated using same data sets but rely on slightly 
different parts of the RGB.  The agreement of the two estimates of
metallicity suggests that the RGB
has been reasonably well determined and that $\feh\ = -1.58 \pm 0.05 $.
Similarly the two estimates of the reddening
yield very good agreement suggesting that 
$ E(B-V) = 1.10 \pm 0.10 $ and thus $ E(V-I) = 1.50 \pm 0.10 $

\subsection{ The Distance Modulus}

Estimates of the distance modulus are hampered by the continuing
controversy over the metallicity dependence of the absolute magnitude of
the HB.  In Table~\ref{tab:prevobs}, for example, two of the relations
which were used to determine the distance to NGC 6517 result in distance
moduli differing by 0.26 magnitudes for the same input parameters.  In
this paper, we will rely on a recent determination which takes into
account revisions in the oxygen abundances and in opacity tables
(\cite{Dorman92a} and \cite{Bergbusch92}).

{}From our differentially corrected \cmd, Figure~\ref{fig:final},
we estimate the
magnitude of the HB to be $V({\rm HB}) = 19.3 \pm 0.10$ which, when combined
with our adopted reddening of $E(B-V)=1.15 \pm 0.10$ and $R_V = 3.8$, yields 
$$V({\rm HB})_0 = 14.9 \pm 0.15.$$
{}From the previous section we adopt a metallicity of $\feh\ = -1.58 \pm 0.05$ for
the cluster, and using \begin{equation} M_V =
0.15\feh\ + 0.83 \end{equation} (\cite{Dorman92a})
derive an absolute magnitude of the
HB of $$M_V({\rm HB}) = 0.60 \pm 0.03.$$ We conclude that the absolute distance
modulus of NGC 6517 is $$(m - M)_0 = 14.3 \pm 0.2,$$ implying that the
cluster is more distant than has heretofore been realized.
We will return to this point in our concluding discussion.

\begin{figure}[hbt]
\hspace*{\fill}{\psfig{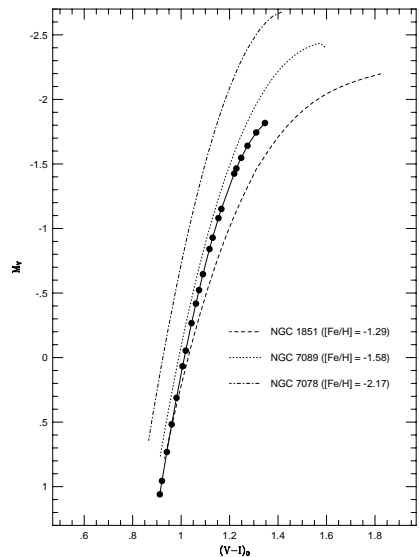}}\hspace*{\fill}
\caption[The Fiducial RGB]{
The giant branch plotted as fiducial points chosen to reflect the
most likely position of the unobscured RGB on the absolute scale.
  Also shown are three RGBs with
comparable metallicities, (\cite{DaCosta90})}
\label{fig:compare}
\end{figure}

\section{Summary and Conclusions}

Our principal conclusions are the following:

\begin{itemize}

\item  NGC 6517 is differentially reddened with an amplitude of
$E(V-I) \sim 0.4$ across our $\sim 4$ arc-min field.  The gradient in
the reddening is spatially consistent with the known clumpiness in the
cloud distribution in the direction of NGC 6517.

\item The mean foreground reddening we derive, $E(B-V) = 1.10 \pm 0.05,$
is consistent with values obtained by other authors in earlier studies.

\item
The absolute distance modulus we find for NGC 6517, $(m-M)_o = 14.3 \pm 0.2,$
is larger than the rough value estimated by Harris (1975):
his {\it apparent} distance modulus in $V$, 18.1, combined with the foreground
absorption of $A_V = 4.0 $ yields a true distance modulus of 14.1.  Of
course, his subsequent downward revision (Harris 1980) by 0.6 mag makes the
difference larger still.  This estimate relied on the level of the tip of
the RGB which may be confused with the large number of foreground
stars (Figure~\ref{fig:final}).

\item
Our determination of the cluster's metallicity, $\feh = -1.58 \pm 0.05,$
confirms that it is a halo
object.  This measurement lies between the two previous determinations and
is indicative of the trouble associated with crowded field spectroscopy.  We
feel that this determination of the metallicity using the \cmd\ gives a more
accurate picture of the general nature of the cluster with good discrimination 
possible between cluster and field stars.
This metallicity is confirmed by a comparison of the
fiducial sequence  of NGC 6517 with clusters
in the same metallicity range (see Figure~\ref{fig:compare})
 and by the blue HB, which is a characteristic of
many metal-poor systems.

\item Our newly-determined distance for NGC 6517 puts it {\em within}
the Galactic bulge ($R_{GC} \sim 3.0$\ kpc for $R_0$ in the range 8.0-10.0\
kpc).  We conclude, then, that NGC 6517 {\em} is a bulge
globular cluster with an intermediate to low metallicity.  

\item The RGB of NGC 6517 is rather stubby. 
 It is the third cluster in the bulge with such a feature 
(\cite{Janes91} and \cite{Stetson94}).
This seems significant as only four bulge objects have \cmds\ of 
sufficient quality to distinguish this feature.
Janes \& Heasley (1991) suggest that the stubby RGB of NGC 6293 may be
attributable to the fact that the cluster has undergone core collapse.
If this is the case then NGC 6517 and NGC 6287 should also be investigated 
for possible core collapse.  On the other hand, this truncated RGB may
be indicative of some other environmental influence within the core
of the Galaxy.

\end{itemize}

{\bf Acknowledgements}

We are pleased to thank the staff at CFHT for their excellent support,
This work was supported in part by an Operating Grants to DAH and WEH
{}from the Natural Sciences and Engineering Research Council of Canada.
JJK expresses his appreciation to Queen's University for financial
support.  The authors also thank Ata Sarajedini for his helpful comments during
the preparation of this paper.

\bibliographystyle{baron}
\bibliography{paper}

\end{document}